\documentclass[letterpaper, 10 pt, conference]{ieeeconf}  

\IEEEoverridecommandlockouts                              

\usepackage{amsmath} 
\usepackage{amssymb}  
\usepackage{graphicx}
\usepackage{multirow}
\usepackage[table]{xcolor}
\usepackage{xcolor}
\usepackage{hyperref}
\usepackage{import}
\usepackage{booktabs}
\usepackage{footmisc}
\usepackage{float}

\title{\LARGE \bf
Automated Multi-Drugs Administration During Total Intravenous Anesthesia Using Multi-Model Predictive Control*}

\author{Bob Aubouin--Pairault$^{1,2}$, Mirko Fiacchini$^{1}$, Thao Dang$^{2}$
\thanks{*This work has been partially supported by the LabEx PERSYVAL-Lab (ANR-11-LABX-0025-01) funded by the French program Investissement d’avenir, and by the French-Japanese ANR-CREST project CyPhAI}
\thanks{$^{1}$ Univ. Grenoble Alpes, CNRS, Grenoble INP, GIPSA-lab, 38000 Grenoble, France 
\noindent
{\tt\small \{bob.aubouin-pairault, mirko.fiacchini\}@gipsa-lab.fr}}%
\thanks{$^{2}$ Univ. Grenoble Alpes, CNRS, Grenoble INP, VERIMAG, 38000 Grenoble, France
\noindent
{\tt\small thao.dang@univ-grenoble-alpes.fr}}%
}

\begin{document}

\maketitle
\thispagestyle{empty}
\pagestyle{empty}

\begin{abstract}

In this paper, a multi-model predictive control approach is used to automate the co-administration of propofol and remifentanil from bispectral index measurement during general anesthesia. To handle the parameter uncertainties in the non-linear output function, multiple Extended Kalman Filters are used to estimate the state of the system in parallel. The best model is chosen using a model-matching criterion and used in a non-linear MPC to compute the next drug rates. The method is compared with a conventional non-linear MPC approach and a PID from the literature. The robustness of the controller is evaluated using Monte-Carlo simulations on a wide population introducing uncertainties in the models. Both simulation setup and controller codes are accessible in open source for further use. Our preliminary results show the potential interest in using a multi-model method to handle parameter uncertainties.
\end{abstract}



\section{Introduction}
The main task of an anesthesiologist during general anesthesia is to monitor and regulate the administration of intravenous drugs to achieve the desired level of hypnosis and analgesia while maintaining stable physiological signals. With the advent of quick-acting intravenous drugs like propofol and remifentanil, and the use of EEG-based hypnotic indicators such as the bispectral index (BIS), researchers have been exploring the possibility of automating the drug delivery process \cite{copotAutomatedDrugDelivery2020}.
\newline

The goal of developing a closed-loop method for administering anesthesia drugs is to improve the patient's state evolution and reduce the workload for anesthesiologists. So far, studies have demonstrated the benefits of using closed-loop control for anesthesia drugs \cite{brogiClinicalPerformanceSafety2017}, \cite{pasinClosedLoopDeliverySystems2017}, but research is ongoing to identify the best and most reliable control method \cite{loebClosedLoopAnesthesiaReady2017}. The task of automating drug dosage during general anesthesia is a complex and ongoing area of research that has been an objective for the control community for over the last two decades. The high level of reliability required, along with the uncertain nature of the system, makes it a difficult task to design a controller. Numerous closed-loop control strategies have been proposed, see surveys \cite{singhArtificialIntelligenceAnesthesia2022} and \cite{ghitaClosedLoopControlAnesthesia2020} for instance. Due to the lack of reliable measurement of the analgesia level, most of the papers focus on the propofol-BIS SISO system, which is the kind of controller most widely clinically tested. However, the dosage paradigm during a real surgery is much more complex, as the anesthesiologist needs to take into account the synergic effect between remifentanil and propofol and the side effect of those drugs on the hemodynamic system. In this paper, the problem of designing a controller for the MISO system propofol-remifentanil to BIS is addressed.
\newline

Multi-model approaches to handle large parameters uncertainties have been well detailed in \cite{andersonOptimalFiltering2012} and \cite{narendraAdaptiveControlUsing1997}, and used for drug control in \cite{raoExperimentalStudiesMultiplemodel2003} to regulate mean arterial pressure and cardiac output in critical care subjects. To the authors' knowledge, this method has not been exploited for the control of the anesthesia process. However, analogous ideas to deal with the model parametric uncertainties for reducing the inter-patient variability for the SISO system propofol-BIS have been recently considered in \cite{wahlquistIndividualizedClosedloopAnesthesia2020}. 

The problem of control design for propofol and remifentanil rates given the BIS measurements has already been studied during the last decade. In \cite{ionescuEvaluationPropofolRemifentanil2011} and \cite{nascuEvaluationThreeProtocols2011} an EPSAC MPC has been designed using a linearized model of drug synergies, and simulations on a small set of patients has shown the superiority of this method compared to an approach with heuristic rules for the injection of remifentanil. In \cite{liuClosedLoopCoadministrationPropofol2011} a dual PID along with a heuristic-based approach has been clinically tested with good performance. Work \cite{nogueiraControllingDepthAnesthesia2014} proposes a positive control law allowing real-time tuning of the propofol-remifentanil balance while ensuring stability. In \cite{padmanabhanReinforcementLearningbasedControl2017}, a Reinforcement Learning method has been used to address the challenge of the MISO system control design with simulation testing. The authors of \cite{westDesignEvaluationClosedLoop2018} put forward a mid-range controller strategy that leverages the use of remifentanil for short-term and small-scale modulation of the bispectral index (BIS), while relying on propofol for longer-term interventions. This idea has been then formalized in \cite{vanheusdenRobustMISOControl2018} and \cite{eskandariExtendedHabituatingModel2020} where an $H_{\infty}$ and an MPC controller have been respectively tested with clinical trials and simulations. More recently in \cite{merigoOptimizedPIDControl2019b}, \cite{schiavoIndividualizedPIDTuning2022a}, and \cite{pawlowskiModelPredictiveControl2022a} the authors have used the idea of fixing the ratio between drug flow rates to propose a PID and an MPC controller. To assess, the robustness of those last controllers, uncertainties have been introduced in the model and Monte-Carlo simulations have been performed.
\newline

In this study, a multi-model state estimation method together with a predictive control strategy is proposed for co-administering propofol and remifentanil using only the BIS as a measured output (MISO system). The method is based on the use of a multi-model parallel implementation of Extended Kalman Filters, followed by a model selection algorithm, and finally, a non-linear MPC to determine the optimal control input based on the selected model. The novelty of this approach resides in a method that can address the uncertainties in the non-linear functions involved in the anesthesia model. This research is intended as a preliminary investigation to demonstrate the feasibility of the control strategy before addressing more complex MIMO systems, where the mean arterial pressure could be used as output for instance. The method is tested on the induction phase which corresponds to the beginning of the anesthesia, when the patient fall asleep. From a control point of view, this is the most challenging part of the anesthesia process since the patient drug reaction is the most uncertain. Compared to the recent literature on the topic, \cite{merigoOptimizedPIDControl2019b}, \cite{schiavoIndividualizedPIDTuning2022a} and \cite{pawlowskiModelPredictiveControl2022a}, the method presented in this paper does not assume a fixed ratio between drug flow, the balance between the drugs is managed through the optimization process. The robustness of the method is tested using Monte-Carlo simulations and results are compared to those obtained by the PID controller presented in \cite{merigoOptimizedPIDControl2019b}.
\newline

The rest of the paper is organized as follows. In Section~2 the standard drug models for anesthesia are recalled along with the associated uncertainties used in the simulations, then in Section~3 the control method is detailed. Finally, Section~4 presents the simulation setup and the associated results. Section~5 provides some concluding remarks.

\section{Standard Anesthesia Model}
Drug models involved in anesthesia dynamics modelling are usually composed of two parts: the Pharmacokinetic (PK) and the Pharmacodynamic (PD). The PK models describe the dynamics of drug concentrations in the patient's body whereas the PD ones represent the link between the drug concentrations and a given physiological effect.

\subsection{Compartment Pharmacokinetic Model}

\begin{figure}[h]
\center
\includegraphics{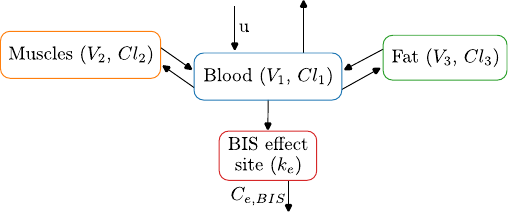}
\caption{Schemes of the PK compartments model}
\label{fig:compartiment}
\end{figure}

For pharmacokinetic (PK) models of both propofol and remifentanil, a common approach is to use a four-compartment model. This model divides the body into three physical compartments: blood, muscles, and fat, and a virtual effect site, as illustrated in Fig.~\ref{fig:compartiment}. The compartments model results in a linear system represented  by the following equations:

\begin{align}\label{eq:PKmodel}
\footnotesize
\begin{pmatrix}
\dot{x}_1 \\ \dot{x}_2 \\ \dot{x}_3\\ \dot{x}_4
\end{pmatrix}
 = & \footnotesize \begin{pmatrix}
 -(k_{10} + k_{12} + k_{13}) & k_{12} & k_{13} & 0\\
 k_{21} & -k_{21} & 0 & 0 \\
  k_{31} & 0 & -k_{31} & 0\\
  k_e & 0 & 0 & -k_e \\
 \end{pmatrix}
 \begin{pmatrix}
x_1 \\ x_2 \\ x_3 \\ x_4
\end{pmatrix} \nonumber \\
&  \footnotesize +  \begin{pmatrix}
\frac{1}{V_1} \\ 0\\ 0 \\ 0
\end{pmatrix} u(t)
\end{align}
\noindent
where $x_1(t),x_2(t)$, $x_3(t)$, and $x_4(t)$ represent respectively the drug concentration in blood, muscle, fat, and effect site. The coefficients can be determined from the equation~(\ref{eq:coeff}), except $k_e$ which is not related to physical meaning.

\begin{equation} \label{eq:coeff}
\begin{aligned}
k_{10} = \frac{Cl_1}{V_1},   k_{12} = \frac{Cl_2}{V_1},   k_{13} = \frac{Cl_3}{V_1} ,\\
k_{21} = \frac{Cl_2}{V_2} , k_{31} = \frac{Cl_3}{V_3}
\end{aligned}
\end{equation}

\noindent
with $V_i$ and $Cl_i$ ($i=1,2,3$) respectively the volume and the clearance rates of each compartment which can be computed from a population-based model as in \cite{schniderInfluenceAgePropofol1999} and \cite{mintoInfluenceAgeGender1997}. The input $u(t)$ is the drug infusion rate. Next, the notation $x_{p}$ and $x_{r}$ for the states of the compartment model for propofol and remifentanil is used. Also, $A_p$, $B_p$, $A_r$, and $B_r$ are the state and input matrix of both drugs. Finally, both compartment models can be described by the decoupled system:
\begin{gather}
    \begin{pmatrix}
        \dot{x}_p \\ \dot{x}_r 
    \end{pmatrix}
    =
    \begin{pmatrix}
        A_p & 0^{4 \times 4 }\\
         0^{4 \times 4} &  A_r
    \end{pmatrix}
        \begin{pmatrix}
        x_p \\ x_r 
    \end{pmatrix}
    + 
        \begin{pmatrix}
        B_p & 0^{4 \times 1 } \\ 
        0^{4 \times 1 }  &   B_r
    \end{pmatrix}
            \begin{pmatrix}
        u_p \\ u_r 
    \end{pmatrix}.
    \label{G_Model}
\end{gather}

\subsection{Pharmacodynamic Model}
The impact of drug concentration on the bispectral index (BIS) is typically modeled using a Hill function. Due to the synergic effect between propofol and remifentanil, the effect can be modeled as a response surface model \cite{mintoResponseSurfaceModel2000}:

\begin{equation}
\label{eq:BIS}
BIS(t) = E_{0}\left(1 - \frac{U(t)^\gamma}{1 + U(t)^\gamma}  \right)
\end{equation}
\noindent
with $E_0$ the initial BIS, $\gamma$ the slope coefficient of the surface and $U(t)$ the interaction term defined by:


\begin{equation}
U(t) = \frac{x_{p4}(t)}{C_{50p}} + \frac{x_{r4}(t)}{C_{50r}}.
\label{eq:U}
\end{equation}


In these equations, $x_{p4}$ and $x_{r4}$ are the propofol and remifentanil concentrations of the BIS effect-site, $C_{50p}$ and $C_{50r}$ are the propofol and remifentanil half-effect concentration for BIS ({\em i.e.} the concentrations to obtain half of the effect of the drugs).

Finally, the fully discretized model subject to noise can be summarized by the following structure:

\begin{equation}
\label{eq:model}
\begin{split}
x(k+1) &= A x(k) + Bu(k)\\
BIS(k) &= h(x(k)) + w(k)
\end{split}
\end{equation}
\noindent
where $h$ is the non-linear output function from eq.~(\ref{eq:BIS})-(\ref{eq:U}) and $w$ is the measurement noise.

\subsection{Model Parameters and Uncertainties}
\label{sec:uncertain}

Several studies have been conducted in order to link the patient characteristics (age, height, weight, sex) to the PK parameters. For control purposes, the most widely accepted are the models developed in \cite{schniderInfluenceMethodAdministration1998a} for propofol and in \cite{mintoInfluenceAgeGender1997} for  remifentanil. To simulate uncertainties in our testing procedure, Monte-Carlo simulations are used with a log-normal distribution for each parameter. The standard deviations used are those given in the papers cited above, nominal values are available in Table~\ref{tab:uncen_PK}.

\begin{table}[H]
\center
\rowcolors{3}{gray!20}{white}
\begin{tabular}{l|c|c|c|c}
 & \multicolumn{2}{c|}{propofol} & \multicolumn{2}{c}{remifentanil} \\
 & nominal & log std  & nominal  & log std \\
 \hline
 $V_1$ 	&	4.27	&	0.17	&	5.22	&	0.26 \\
 $V_2$ 	&	25.94	&	0.25	&	10.26	&	0.28 \\
 $V_3$ 	&	238	    &	2.66	&	5.42	&	0.60 \\
 $Cl_1$ &	1.64	&	0.16	&	2.69	&	0.14 \\
 $Cl_2$ &	1.72	&	0.02	&	2.20	&	0.35 \\
 $Cl_3$ &	0.84	&	0.10	&	0.08	&	0.39 \\
 $k_e$ 	&	0.456	&	0.19	&	0.63	&	0.62 \\
\end{tabular}
\caption{Nominal values are for a man of 70kg, 170cm, and 35 years old. log std stands for logarithmic standard deviation.}
\label{tab:uncen_PK}
\end{table}

For the response surface model, the values from \cite{bouillonPharmacodynamicInteractionPropofol2004} were used, as outlined in Table~\ref{tab:uncen_PD}.

\begin{table}[H]
\center
\rowcolors{2}{white}{gray!20}
\begin{tabular}{l|c|c}
 & nominal & log std \\
 \hline
 $C_{50p}$ 	&	4.47	&	0.18	\\
 $C_{50r}$ 	&	19.3	&	0.76	\\
 $\gamma$ 	&	1.43	&	0.30	\\
 $E_0$ 		&	97.4	&	0		\\
\end{tabular}
\caption{Parameters of the log-normal distribution for the PD parameters}
\label{tab:uncen_PD}
\end{table}

\section{Multi-Model Control}
As previously discussed, drug models are characterized by parameters that might vary significantly from patient to patient. It is then necessary to identify such parameters to improve the control performances, mostly when using controllers strongly relying on the model, as for model predictive control employed here. To address this issue, we propose in this section a multi-model approach. Given that the impact of PD variability uncertainty is more significant than PK variability uncertainty \cite{kriegerModelingAnalysisIndividualized2014a}, the uncertain parameters of the PD system can be considered unknown, and are represented by the vector $\theta = \begin{pmatrix} C_{50p} & C_{50r} & \gamma \end{pmatrix}$. A method is presented to estimate these parameters using data available in the first part of the surgical operation.

\begin{figure}
\center
\includegraphics[width=\columnwidth]{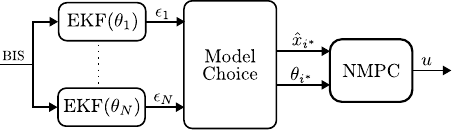}
\caption{Scheme of the MMPC strategy}
\label{fig:MM_schem}
\end{figure}

The multi-model approach consists of three parts, as depicted in Fig.~\ref{fig:MM_schem}. First, the states of the PK models are estimated in parallel using a set of EKFs, one for every realization of the vector selected within a grid in the space of the parameters. The grid is designed to reasonably represent the variability of the parameter vector. Next, a vector is chosen using a model-matching criterion and, finally, a non-linear Model Predictive Controller is used to compute the control input to apply.

\subsection{Extended Kalman Filter}
\label{seq:EKF}
In this section, the basics of the EKF are recalled. EKF is a state estimation method that relies on the linearization of a non-linear model. If we consider the model given in (\ref{eq:model}) with the non-linear function $h$ parametrized by $\theta$, the estimator using the parameter vector $\theta_i$ is given by:

\begin{flalign*}
&H_i(k) = \left. \frac{\partial h(x, \theta_i)}{\partial x} \right| _{x=\hat{x}_i(k_{|k-1})} \\
&K_i(k) = P_i(k_{|k-1})H_i^\top (k)(H_i(k)P_i(k_{|k-1})H_i^\top (k) + R_2)^{-1} \\
&\hat{x}_i(k_{|k}) = \hat{x}_i(k_{|k-1}) + K_i(k)(y(k) - h(\hat{x}_i(k_{|k-1}),\theta_i )) \\
&P_i(k_{|k}) = P_i(k_{|k-1}) - K_i(k) H_i(k) P_i(k_{|k-1}) \\
&\hat{x}_i(k+1_{|k}) =  A \hat{x}_i(k_{|k}) + Bu(k) \\
&P_i(k+1_{|k}) = A P_i(k_{|k})A^\top + R_1
\end{flalign*}.

Here the notation $X(k +1{|k})$, respectively $X(k{|k})$ and $X(k{|k-1})$, represents the value of variable X computed at time step $k+1$ based on the knowledge available at $k$. The estimated state vector is $\hat{x}$ and $P$ is the covariance matrix. $R_1$ and $R_2$ are two constant matrices used to respectively characterize the process perturbations and the measurements noise.

Since negative concentrations are not allowed, a saturation is added to the state estimation expression after the measurement update: 

\begin{equation*}
\hat{x}(k_{|k}) = \max(0, \, \hat{x}(k_{|k})).
\end{equation*}

\subsection{Model Selection}

To choose a model to predict the state trajectory of the system in the future, the model-matching criterion proposed in \cite{narendraAdaptiveControlUsing1997} is used. In order to determine the model matching criterion $J_i$ for the $i^{th}$ estimator at time $k$, the following method is used: 
\begin{itemize}
    \item Each Extended Kalman Filter (EKF) generates a state estimate, which is stored;
    \item A state trajectory $x(l)$ for $l\in \{k-N_c,..k\}$ is formed by using equation \eqref{eq:model} with the initial point set to the estimation value previously saved at the time stamp $k-N_c$. $N_c$ is the number of samples in the observation window;
    \item The resulting trajectory is then compared to the BIS measurement to obtain the prediction error $\epsilon_i(l) = h(x(l), \theta_i) - y(l)$.
     \item Finally, the criterion is computed using the following formula:
     \begin{equation}
        J_i(k) = \alpha \epsilon_i^2(k) + \beta \sum^{N_c}_{l=0} e^{-\lambda l}\epsilon^2_i(k-l)
    \end{equation}
     where $\alpha$, $\beta$, and $\lambda$ are three positive constants used to tune the convergence rate.
\end{itemize}

To circumvent potential instability in the model selection process, a threshold is employed to identify the optimal model. Specifically, the estimator with the smaller criterion is selected if the difference between its criterion and the criterion of the previously chosen estimator exceeds a predetermined threshold. 

\subsection{Model Predictive Control}
\label{seq:MPC}
Model Predictive Control is an advanced control method that uses online optimization to obtain optimal control input in presence of constraints on the state and the control input \cite{rawlingsjamesb.ModelPredictiveControl2009}.  In this paper, a non-linear MPC parametrized by the parameter vector $\theta$ is used.
The cost of the optimization problem is given by:

\begin{equation}
\begin{aligned}
J = & \sum_{i=1}^{N}(y_{ref}(k)-h(x(k+i),\theta))^2\\
& + \sum_{i=1}^{N_u}(u(k+i))^T R  (u(k+i))^2.
\end{aligned}
\end{equation}

%
%
%
%
%
%

The reference signal to be followed is denoted $y_{ref}$, while the associated control input values are subject to a cost matrix $R$, which enables the user to modulate the balance between propofol and remifentanil. $N$ and $N_u$ are respectively the prediction and the control horizon. As specified in \cite{merigoOptimizedPIDControl2019b}, the maximum infusion rate of propofol is $6.67 mg/s$ and that of remifentanil is  $16.67 \mu g/s$. Thus, an optimization problem with constraints given by the system dynamics and the bounds on the inputs is obtained. Note that a quartic cost is utilized to achieve a better trade-off between undershooting and rapidity.


 
In order to ensure the convergence of the system at the desired BIS target despite the presence of uncertainties and disturbances, an integrator is added to the MPC internal reference after the induction phase (after 2 minutes in practice):
\begin{equation*}
y_{ref}(k+1) = y_{ref}(k) + k_i (BIS_{target} - BIS(k))
\end{equation*}
\noindent
where $k_i$ is a constant used to tune the convergence speed.

\section{Numerical Simulations}

In this section, three controllers are tested on the same induction scenario to obtain a fair comparison. First, the PID from \cite{merigoOptimizedPIDControl2019b}, then an EKF estimator associated with a non-linear MPC (NMPC) as described in the sections~\ref{seq:EKF} and \ref{seq:MPC} with the nominal parameter vector $\theta$ and, finally, the proposed multi-model predictive controller (MMPC) previously detailed.

\subsection{Controller tuning}

The PID was tuned using a particle swarm optimization over a randomly sampled patient cohort drawn from the distribution detailed in Table~\ref{tab:uncen_PD} as in \cite{merigoOptimizedPIDControl2019b}. The ratio between propofol and remifentanil rates was set to $2$ as this is a common value. Additionally, the system's sampling time was set to $1$ second, following the protocol outlined in the same paper.

For the NMPC and the MMPC, the parameters were tuned by hand on the same patient table. The EKF was tuned first to ensure the convergence of the state estimation. Secondly, the prediction horizon of the MPC has been set to $1$ minute with a sampling time of $2$ seconds, then the cost matrix $R$ was set to get the ratio between drugs and the response time analogous to those of PID. Finally, for the NMPC, the integrator constant $k_i$ was selected to achieve a balance between fast convergence of the system and reduced oscillations.

For the MMPC, the distribution of the parameters $\theta_i$ was tuned to obtain good performance for each real model, even for the more sensitive ones, that are those with smaller $C_{50}$. For the simulations done in this paper, 45 models have been used in parallel with a grid distribution across the three parameters. To adjust the parameters of the model selector, the speed at which the selection converges during the induction phase was used as an indicator. The window length was chosen to be the same as the one of the MPC ($N_c=30$) with $\alpha=0$, $\beta=1$, $\lambda=0.05$, and $\delta = 30$.

\subsection{Simulation Setup}

\begin{table*}
\center
\rowcolors{3}{white}{gray!20}
\begin{tabular}{l|cc|cc|cc|cc|cc}
\toprule
\multirow{2}{*}{Controller} & \multicolumn{2}{c|}{TT (min)} & \multicolumn{2}{c|}{BIS\_NADIR} & \multicolumn{2}{c|}{ST10 (min)} & \multicolumn{2}{c|}{ST20 (min)} & \multicolumn{2}{c}{US}\\

	    & mean $\pm$ std & max & mean $\pm$ std & min & mean $\pm$ std & max & mean $\pm$ std & max & mean $\pm$ std & max\\ 
\midrule
PID & \textbf{1.21$\pm$0.25} & \textbf{2.13} & 46.07$\pm$3.21 & 24.2 & \textbf{1.66$\pm$0.97} & 8.95 & \textbf{1.13$\pm$0.51} & 5.43 & 0.76$\pm$2.25 & 20.8 \\
NMPC & 2.43$\pm$1.27 & 7.67 & 48.16$\pm$2.48 & 34.44 & 2.58$\pm$1.21 & \textbf{7.67} & 1.88$\pm$0.82 & 5.9 & 0.29$\pm$1.25 & 10.56 \\
MMPC & 2.22$\pm$0.94 & 9.73 & \textbf{48.37$\pm$1.69} & \textbf{38.05} & 2.32$\pm$1.09 & 9.73 & 1.76$\pm$0.52 & \textbf{5.03} & \textbf{0.07$\pm$0.47} & \textbf{6.95} \\
\bottomrule
\end{tabular}
\caption{Performance criteria for the three controllers in the induction phase}
\label{tab:results}
\end{table*}

To assess the performances of the controllers, simulations are done with 500 different patients using random uniform sampling to obtain age, sex, height, and weight. Then uncertainties are added to both the PK and PD models with log-normal sampling as described in Section~\ref{sec:uncertain}. This simulation is used to test the controllers on a wide range of patient profiles. The performance criteria are those proposed in \cite{ionescuRobustPredictiveControl2008} and also used in \cite{merigoOptimizedPIDControl2019b}, \cite{schiavoIndividualizedPIDTuning2022a}, and \cite{pawlowskiModelPredictiveControl2022a}. They are listed below:
\begin{itemize}
	\item \textit{Time to target} (TT): time to reach the target BIS interval [45, 55].
	\item \textit{BIS NADIR}: minimum BIS value reached during the induction phase.
	\item \textit{Settling time 10} (ST10), respectively (ST20): time to reach the interval target $\pm 10\%$, respectively $20\%$, and stay within this range.
	\item \textit{Undershoot} (US): maximum undershoot below a BIS of 45 during the induction phase.
\end{itemize}

The involved optimization problems are solved using CASADI software \cite{anderssonCasADiSoftwareFramework2019} with IPOPT solver. The maximum computation time of the proposed solution for one step is 0.14s, which makes it a plausible solution.
The whole code to perform the simulations presented in the paper is written in Python and available at \url{https://github.com/BobAubouin/TIVA_Drug_Control} and uses \cite{aubouin-pairaultPASPythonAnesthesia2023} to perform all the simulations. Note that the control design method presented in this paper is the second open-source controller for anesthesia (after the one presented in \cite{comanAnesthesiaGUIDEMATLABTool2022}), and it has been shared with the hope that this will lead to more easily reproducible results in the future.

\subsection{Results}

\begin{figure}[H]
\center
\includegraphics[width=\columnwidth]{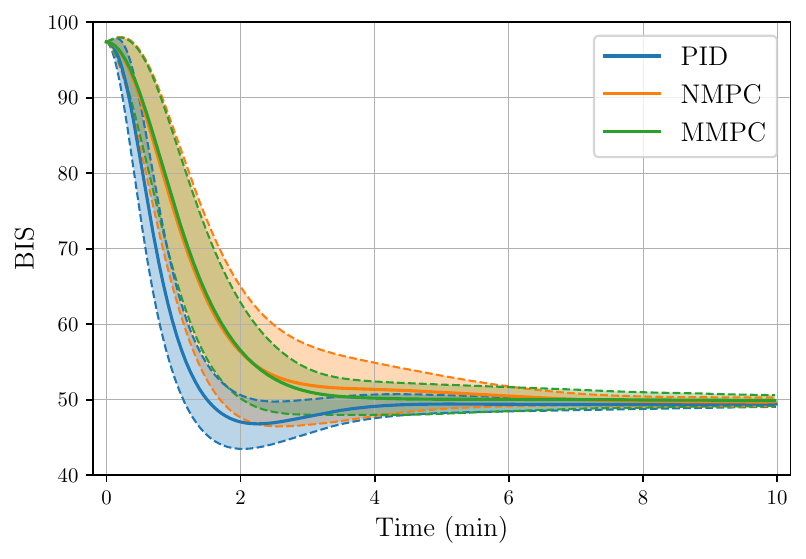}
\caption{Mean BIS over the 500 patients for the three controllers. The plot is the mean value $\pm$ standard deviation.}
\label{fig:mean_BIS}
\end{figure}

\begin{figure}[H]
\center
\includegraphics[width=\columnwidth]{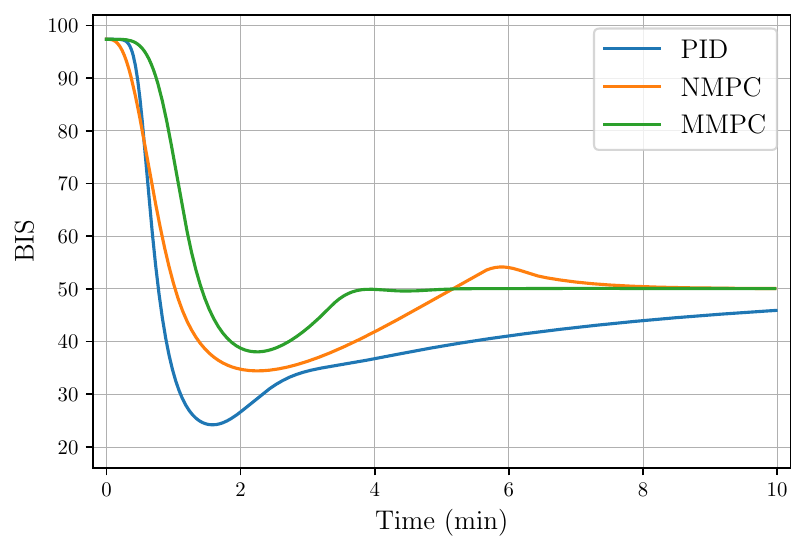}
\caption{BIS values for the worst case of each controller (in terms of undershoot) for the three controllers.}
\label{fig:worst_BIS}
\end{figure}

\begin{figure}[H]
\center
\includegraphics[width=\columnwidth]{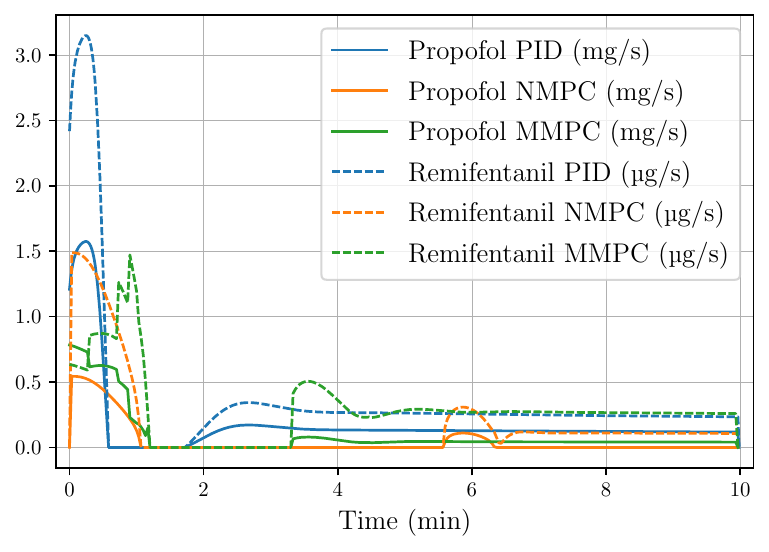}
\caption{Drug rates for the worst case of each controller (in terms of undershoot) for the three controllers.}
\label{fig:worst_U}
\end{figure}

The simulation results are presented in Table.~\ref{tab:results} and Fig.~\ref{fig:mean_BIS}. Moreover, the BIS trajectories of the case with the worst undershoot for each controller are shown in Fig.~\ref{fig:worst_BIS} and the associated control input in Fig.~\ref{fig:worst_U}.

The results of the study indicate that the MPCs exhibit superior performance compared to the PID controllers. Although the PID controllers exhibit faster time to target (TT), this results in a larger undershoot (US) and comparable settling times (ST10 and ST20). On average, the two MPC controllers were found to be comparable, however, the multi-model approach demonstrates its usefulness by reducing the effects of the uncertainties on the output trajectories, as depicted in Fig.~\ref{fig:mean_BIS}. Furthermore, this approach shows huge improvements in cases of patients with extreme models as shown by the maximal undershoot values and Fig.~\ref{fig:worst_BIS}. These results demonstrate the effectiveness of the multi-model approach in quickly identifying the appropriate PD model for dosing propofol and remifentanil in this scenario.

Nevertheless, the conclusion can be mitigated considering that the PID used for the comparison has been optimized on a patient table while in the more recent paper \cite{schiavoIndividualizedPIDTuning2022a} the authors proposed to optimize the PID for each patient characteristic, and thus for each PK model. Since this controller was longer to implement and to test, though, the one from \cite{merigoOptimizedPIDControl2019b} has been used. In \cite{schiavoIndividualizedPIDTuning2022a} the authors explain that this individualized approach allows the controller to reduce undershoot. However similar conclusions should be obtained with the more recent PID version since it is inherent to this architecture which does not handle the PD uncertainties.



\section{Conclusion}
In this paper, a new control method for the co-administration of propofol and remifentanil driven by BIS measurement has been proposed. A multi-model predictive controller has been designed and compared to a PID controller and a non-linear MPC using the average model. The simulations done on a random database of $500$ patients including a high level of uncertainties show the benefit of the multi-model approach for the induction phase.\\
 In the future, the authors will continue this work to assess the controller performance during the maintenance phase and in a noisy environment. It seems that the multi-model approach, which provides the controller with the possibility of learning about the system, can also be effective in rejecting disturbances, however, a noisy environment can slow down the identification process and thus mitigate the results. The final end-point is to extend it to the whole anesthesia regulation problem which includes other system outputs such as hemodynamic signals and analgesic indicators.

\bibliographystyle{ieeetr}
\bibliography{bibli}

\end{document}